\documentclass[preprint,aps]{revtex4}

\usepackage{graphicx}% Include figure files
\usepackage{dcolumn}% Align table columns on decimal point
\usepackage{bm}% bold math
\usepackage{latexsym}
\usepackage{amsfonts}
\usepackage{amssymb}
\usepackage{amsmath}
\usepackage[usenames]{color}
\begin{document}
\preprint{IPMU 08-0036}
\preprint{OCU-PHYS 300}
\preprint{AP-GR 59}
\preprint{KUNS 2143}

\title{Anisotropic Inflation from Vector Impurity}

\author{Sugumi Kanno$^{1)}$}
\author{Masashi Kimura$^{2)}$}
\author{Jiro Soda$^{3)}$}
\author{Shuichiro Yokoyama$^{4)}$}
\affiliation{1) Institute for the Physics and Mathematics
of the Universe, University of Tokyo, Chiba, 277-8568, Japan}
\affiliation{2)Department of Mathematics and Physics,
Graduate School of Science, Osaka City University, Osaka, 558-8585, Japan}
\affiliation{3)Department of Physics,  Kyoto University, Kyoto, 606-8501, Japan
}
\affiliation{4)Department of Physics and Astrophysics, Nagoya University, 
Aichi, 464-8602, Japan}

\date{\today}% It is always \today, today,

%===============================================================%
%************************* ABSTRACT ****************************%
%===============================================================%
\begin{abstract}
We study an inflationary scenario with a vector impurity.
We show that the universe undergoes anisotropic inflationary
expansion due to a preferred direction determined by the vector.
Using the slow-roll approximation, we find a formula to 
determine anisotropy of the inflationary universe.
We discuss possible observable predictions of this scenario.
In particular, it is stressed that primordial gravitational
waves can be induced from curvature perturbations.
Hence, even in low scale inflation, a sizable
amount of primordial gravitational waves may be produced
during inflation.  
\end{abstract}

\maketitle

%===============================================================%
%************************ SECTION I ****************************%
%===============================================================%
\section{Introduction}

It is often mentioned that cosmology has entered into a new stage,
so-called precision cosmology. Of course, it is referring to developments
of observational side.  From theoretical point of view, however,
we have not yet exhausted possible phenomenology on the order of
a few percent. Clearly, it is important to explore qualitatively new
scenarios at the percent level. Here, above all, we would like to 
point out that an inflationary model with a few percent of anisotropy 
yields significant consequences.  Indeed, a few percent does not mean
  the consequent effects are negligible. Rather, 
it provides the leading component of the primordial gravitational waves
in low scale inflationary models which are preferred by recent
 model construction in string theory~
\cite{Kallosh:2004yh}. 

 One may feel that to seek the anisotropic inflation is against for
 the basics of the inflationary scenario.  Actually, 
the isotropy is the most robust property of inflation
 because of the cosmic no-hair theorem
on the isotropization of Bianchi universes~\cite{Wald:1983ky}. 
However, it is possible to evade the cosmic no-hair theorem  by
incorporating the Kalb-Ramond action~\cite{Kaloper:1991rw} or considering
 higher curvature theories of gravity~\cite{Kawai:1998bn,Barrow:2005qv}.  
In spite of the possibility, no one has attempted to construct any inflation 
models based on these ideas. The apparent other possibility is to break
the Lorentz invariance by introducing a condensation of vector field.
The model proposed in~\cite{Ackerman:2007nb,Dulaney:2008bp} 
seems to be successful, however, it is known to be
metastable~\cite{Clayton:2001vy,Dulaney:2008ph,Bluhm:2008yt}. 
A more natural possibility would be to realize a slow-roll phase
of vector fields like as inflaton fields in chaotic inflationary
scenarios. So far, it has been believed that it is difficult to
make the vector field slow-roll without fine tuning~\cite{Ford:1989me}
\footnote{Recently, a non-standard spinor driven inflation has
been proposed~\cite{Boehmer:2007ut}
. It is interesting to see if the
natural slow-rolling can be achieved in this case.}. 
Very recently, the situation has changed by the discovery
of slow-roll mechanism for the vector field 
due to non-minimally coupling~\cite{Golovnev:2008cf}.
Hence, an apparent difficulty to construct the anisotropic inflationary
scenario has been resolved.
At this point, it is important to realize that both scalar and vector fields
are ingredients of
 fundamental particle physics models. 
Therefore, it is natural to consider both scalar and vector 
fields exist during inflation. 
Of course, the vector fields should be subdominant in the 
inflationary dynamics in order to reconcile the 
scenario with current observational data ~\cite{Pullen:2007tu}. 
 In this sense, the vector field should be regarded as impurity.
 Nevertheless, the effect of the vector impurity on observables
should not be overlooked under the current precision cosmology.

In this paper, we propose an anisotropic inflation model 
with the vector impurity. 
To the best of our knowledge, this is the first concrete model which realizes
anisotropic inflation, exits successfully to the isotropic standard universe,
and provides a framework to discuss interesting phenomenology. 
We argue that
the anisotropic inflation yields the statistical anisotropy in fluctuations.
More importantly, as is expected, the primordial gravitational waves could be 
induced from curvature perturbations through the anisotropic background. 
Hence, we can expect the correlation between the curvature perturbations
and the gravitational waves. 
In addition to these, we point out that linear polarization
of the gravitational waves is created, which should be observed through
CMB or direct interferometer observations. 

The organization of this paper is as follows.
In section II, we present the model for the anisotropic inflation.
 In section III, we analyze the system numerically and show that the 
 anisotropic inflation is realized successfully.
  In section IV, using the slow-roll approximation,
we obtain degrees of the anisotropy.
In section V, we discuss phenomenology of the anisotropic inflation
induced by the vector impurity. There, we emphasize that the 
gravitational waves can be produced through the anisotropy of the
spacetime. The final section is devoted to the discussion. 

%===============================================================%
%************************ SECTION II ****************************%
%===============================================================%
\section{The Model}

In this section, we derive the basic equations for the anisotropic
 inflationary scenario induced by the vector impurity.
 The set up is similar to the previous works~
\cite{Dimopoulos:2008rf, Yokoyama:2008xw} where
 scalar and  vector fields are considered.
However, backreaction of the vector field on the background
dynamics is neglected there.
Here, we consider the backreaction and find it
makes significant differences. 

We consider the following action for the background
gravitational field, the scalar field $\phi$ and the
non-minimally coupled massive vector field $A_\mu$:
\begin{eqnarray}
S=\int d^4x\sqrt{-g}\left[~\frac{1}{2\kappa^2}R
-\frac{1}{2}\left(\partial_\mu\phi\right)^2-V(\phi)
-\frac{1}{4}F_{\mu\nu}F^{\mu\nu}  
-\frac{1}{2}\left(m^2-\frac{R}{6}\right)A_\mu A^\mu
~\right] \ ,
\label{action1}
\end{eqnarray}
where $g$ is the determinant of the metric, $R$ is the
Ricci scalar, $V(\phi)$ is the scalar potential, $m$ is 
the mass of the vector field, and 
we have defined  
$F_{\mu\nu}=\partial_\mu A_\nu -\partial_\nu A_\mu$.
The equation of motion for $A_\mu$ from above action:
\begin{eqnarray}
\frac{1}{\sqrt{-g}}\partial_\mu
\left(~\sqrt{-g}F^{\mu\nu}~\right)
=\left(m^2-\frac{R}{6}\right)A^\nu
\end{eqnarray}
reduces $A_0(t)=0$ in the case of $\nu=0$ because of antisymmetry 
of $F^{\mu\nu}$. As we find the vector field has only
spatial components, we take $x$-axis in the direction of the vector,
\begin{eqnarray}
A_\mu=(~0,~A_x(t),~0,~0~),\qquad
\phi=\phi(t) \, .
\label{vector}
\end{eqnarray}

Now, we will take the metric to be homogeneous but anisotropic
Bianchi type-I,
i.e.
\begin{eqnarray}
ds^2=-{\cal N}(t)^2dt^2+e^{2\alpha(t)}\left[~ 
e^{-4\sigma_{+}(t)}dx^2    
+e^{2\sigma_{+}(t)}\left(e^{2\sqrt{3}\sigma_{-}(t)} dy^2 
+e^{-2\sqrt{3}\sigma_{-}(t)}dz^2\right)~\right]\,,
\end{eqnarray}
where ${\cal N}(t)$ is the lapse function. 
With this ansatz, the background action becomes
\begin{eqnarray}
S &=& \int d^4x\frac{1}{{\cal N}}e^{3\alpha}\left[~
\frac{3}{\kappa^2}\left(
-\dot{\alpha}^2+\dot{\sigma}_+^2+\dot{\sigma}_-^2
\right)
+\frac{1}{2}\dot{\phi}^2-{\cal N}^2 V   \right. \nonumber \\
&&  \left.  +\frac{1}{2}\left(\dot{X}-2\dot{\sigma}_+X\right)^2
-\frac{m^2}{2}{\cal N}^2 X^2
+\left(~\frac{1}{2}\dot{\sigma}_+^2
+\frac{1}{2}\dot{\sigma}_-^2
-2\dot{\alpha}\dot{\sigma}_+\right)X^2
~\right]\,,
\label{action2}
\end{eqnarray}
where we have introduced a new variable $X=\exp(-\alpha+2\sigma_+)A_x$
and defined $\cdot=\partial_t$. In general,
this procedure leads to incorrect results. However, for homogeneous
Bianchi models, this reduction procedure is valid as can be checked directly.

The variational equations of motion with respect to ${\cal N}$, $\phi$,
$X$, $\sigma_-$, $\sigma_+$ and $\alpha$ then become (after setting
${\cal N}=1$):
\begin{eqnarray}
&&\frac{3}{\kappa^2}\left(
-\dot{\alpha}^2+\dot{\sigma}_+^2+\dot{\sigma}_-^2
\right)
+\frac{1}{2}\dot{\phi}^2+V
\nonumber \\
&& \qquad
+\frac{1}{2}\left(\dot{X}-2\dot{\sigma}_+X\right)^2
+\frac{m^2}{2}X^2  
 +\left(~\frac{1}{2}\dot{\sigma}_+^2
+\frac{1}{2}\dot{\sigma}_-^2
-2\dot{\alpha}\dot{\sigma}_+\right)X^2
=0\,,\label{eq:N}\\
&&\ddot{\phi}+3\dot{\alpha}\dot{\phi}+V_{,\phi}=0\,,
\label{eq:phi}\\
&&\ddot{X}+3\dot{\alpha}\dot{X}
+\left(m^2-2\ddot{\sigma}_+-2\dot{\alpha}\dot{\sigma}_+
-5\dot{\sigma}_+^2-\dot{\sigma}_-^2
\right)X=0\,,
\label{eq:X}\\
&&\left[~e^{3\alpha}\left(
\frac{6}{\kappa^2}+X^2\right)\dot{\sigma}_-
~\right]^{\cdot}=0\,,
\label{eq:sigma-}\\
&&\left[~e^{3\alpha}\Biggl\{\left(
\frac{6}{\kappa^2}+5X^2
\right)\dot{\sigma}_+-2X\dot{X}-2\dot{\alpha}X^2
\Biggr\}~\right]^\cdot
=0\,,
\label{eq:sigma+}\\
&&\ddot{\alpha}+3\dot{\alpha}^2-\kappa^2V
+\frac{2}{3}\kappa^2\dot{\sigma}_+X\dot{X}
+\kappa^2\left(\frac{1}{3}\ddot{\sigma}_+
+\dot{\alpha}\dot{\sigma}_+-\frac{1}{2}m^2
\right)X^2=0\,.
\label{eq:alpha}
\end{eqnarray}
where $V_{,\phi}\equiv\frac{dV}{d\phi}$.
Notice that the effective mass squared for $X$, i.e. 
$m^2-2\ddot{\sigma}_+-2\dot{\alpha}\dot{\sigma}_+
-5\dot{\sigma}_+^2-\dot{\sigma}_-^2$ in Eq.~(\ref{eq:X})
might look different from the mass term of $X$ in the action
(\ref{action2}). However, we can identify them by transforming
the action (\ref{action2}) into the canonical form for $X$. 

Note also that if the effective mass squared of $X$ in Eq.~(\ref{eq:X}) 
is negative, the system will be tachyonic. 
To avoid this, we require 
the effective mass squared is positive,
$m^2-2\ddot{\sigma}_+-2\dot{\alpha}\dot{\sigma}_+
-5\dot{\sigma}_+^2-\dot{\sigma}_-^2 >0$,
that is, 
$m\neq 0$
 and $\dot{\sigma}_{\pm}$ has to be 
sufficiently small. As we will see later in Eq.~(\ref{eq:sigma+2}), 
this 
latter condition implies the amplitude of the vector field $X$ should be small.
In this sense, the vector field is a kind of impurity. 
In the limit, $\kappa X \rightarrow 0$,
the above set of equations reduce to those of conventional
inflation scenarios. Also, taking look at Eqs.(\ref{eq:N})-(\ref{eq:alpha}),
 we see the deviation from the conventional slow-roll dynamics
comes in of the order $X^2$. 

 Although we have assumed the direction of the vector field does not change
in time and considered only $x$-component of it 
for simplicity in Eq.~(\ref{vector}), we could also incorporate 
other spatial components of the vector field.
  In that case, each component of the vector field 
 obeys the similar equation as $X$. The vector field
turn out to start to rotate under arbitrary initial conditions.  
 If the slow-roll conditions are satisfied for each, the vector will 
 spiral down slowly.
 It is interesting to study the consequence of this complicated dynamics.
 However, we leave it for future work.

\section{Anisotropic Inflation}

In this section, we consider the evolution of the background spacetime
driven by the scalar field $\phi$, which we refer to as inflaton below, 
in the presence of the vector impurity. 

Let us consider the universe after a sufficient expansion, 
$\alpha\rightarrow\infty$.
It is straightforward in this limit to integrate Eqs.~(\ref{eq:sigma-})
and (\ref{eq:sigma+}) to find
\begin{eqnarray}
\dot{\sigma}_-=0\,,\qquad
\dot{\sigma}_+=\frac{1}{6/\kappa^2+5X^2}\left(
2X\dot{X}+2\dot{\alpha}X^2
\right)\,.
\label{attractor}
\end{eqnarray} 
We find that the anisotropy in $y$-$z$ plane, $\dot{\sigma}_-$, will disappear.
This is reminiscent of the comic no-hair theorem
by Wald~\cite{Wald:1983ky}.
However, as long as the vector field exists $X\neq 0$, the anisotropy in $x$ 
direction, $\dot{\sigma}_+$, will still remain even if the universe undergoes 
a period of inflation. In the proof of the cosmic no-hair theorem for
Bianchi models, the strong and dominant energy conditions
are assumed. The reason why the anisotropy does not disappear in our model is 
that the non-minimal coupling breaks the 
strong energy condition.
Thus, we can restrict our metric to the following form:
\begin{eqnarray}
ds^2=-
dt^2
+e^{2\alpha(t)}\left[~ 
e^{-4\sigma_{+}(t)}dx^2    
+e^{2\sigma_{+}(t)}\left( dy^2 + dz^2\right)~\right]\,.
\label{metric}
\end{eqnarray}
The above result might be generalized to
the modified version of the cosmic no-hair theorem.
It is interesting to examine other Bianchi type models in the presence
of the vector impurity. In particular, it is intriguing to prove the cosmic
no-hair theorem for these cases. 

\begin{figure}
\includegraphics[height=8cm, width=12.5cm]{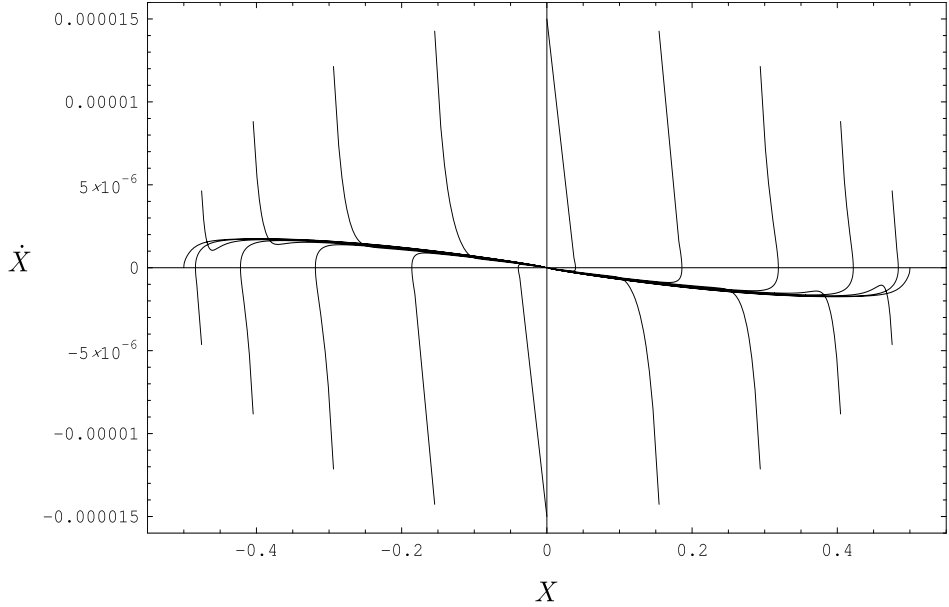}
\caption{The phase flow in $X$-$\dot{X}$ plane is depicted.
For various initial conditions with small amplitude of $X$, we have
plotted the trajectories. Every trajectory converges to the slow roll 
attractor. }
\label{fig:1-a}
\end{figure}

\begin{figure}
\includegraphics[height=8cm, width=13cm]{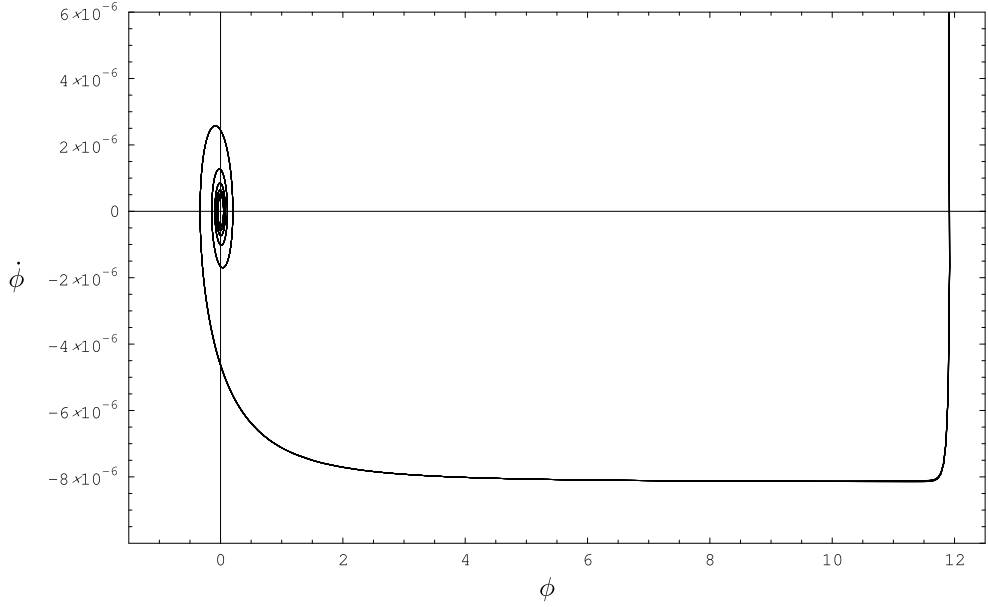}
\caption{The phase flow in $\phi$-$\dot{\phi}$ plane is depicted. 
For the same initial conditions as Fig.1, we have plotted the 
trajectories. Every trajectory is degenerated irrespective of the
initial conditions for X.
As we can see, the inflaton rolls down the potential slowly. The 
inflaton begins to oscillate around
the minimum of the potential and the inflation ends with the reheating.
Here we have plotted a part of the graph for the sake of visualization
, though we started from $\phi_0=10$.}
\label{fig:1-b}
\end{figure}

Now, we numerically solve Eqs.~(\ref{eq:N})-(\ref{eq:alpha}) by setting $\sigma_{-} =0$. 
We will use new variables $H=\dot{\alpha}$ and
 $\Sigma=\dot{\sigma}_+$ below.
We take $V= 1/2 \mu^2 \phi^2$ as the potential for the 
inflaton $\phi$. The parameter of the system is the ratio 
$m/\mu$.
For this calculation, we set $\kappa=1$, $\mu = 10^{-5}$ and $m=2\sqrt{2}\times10^{-5}$.
And, we took the initial values $\phi_0 =10$ and $H_0 = 3\times 10^{-4}$
for all figures.

\begin{figure}
\includegraphics[height=8cm, width=13cm]{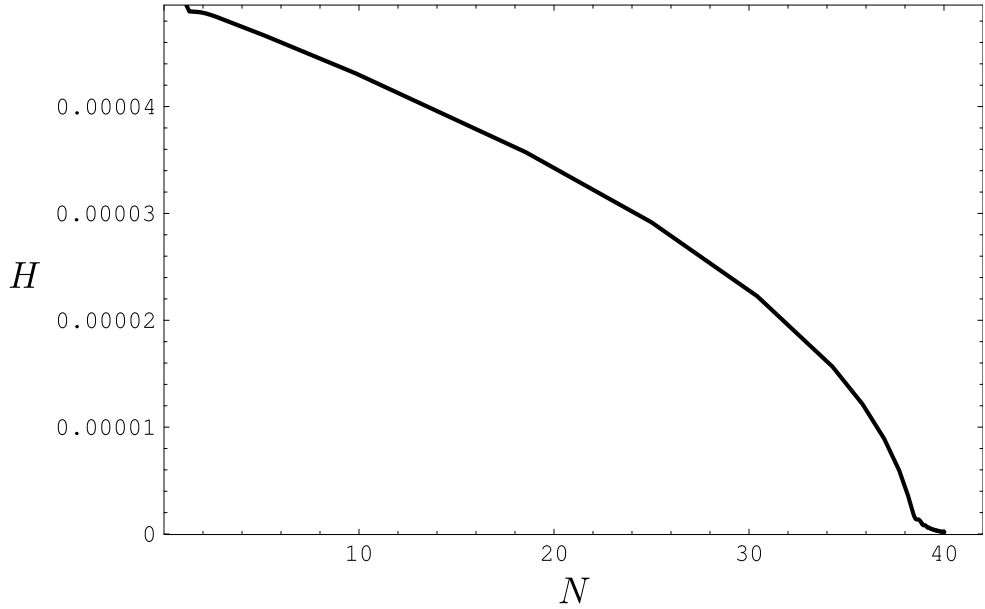}
\caption{The Hubble parameter $H$ is plotted as a function of 
e-folding number $N$.
We see a slow-roll phase clearly. We have e-folding number
$N\sim$ 40 for this case.}
\label{fig:2-a}
\end{figure}

In Fig.~\ref{fig:1-a}, we depicted the phase flow in $X$-$\dot{X}$ plane.
We set the initial value $\Sigma_0 =0$  
and determined $\dot{\phi}_0$ using the constraint
equation (\ref{eq:N}).
We see that slow-roll phase of the vector field is an attractor
when the amplitude of the vector field is sufficiently small.
For the appropriate parameter $m/\mu$,
 the small value such as $X^2 \sim 0.1$
is easily attainable. 
Both the mass and the non-minimal coupling
play important roles for realizing the slow-roll. They affect
the coefficient of $X$ in Eq.~(8). In detail, the mass is important for the
system to be stable and the non-minimal coupling cancels out the term 
$\dot{\alpha}^2$ in the effective mass squared of $X$ which enable the vector
field to slow-roll. Thus, both mass and
non-minimal coupling help the vector field to slow-roll.
Thanks to the slow-roll for the vector field, we have enough anisotropic era.
After the vector field goes to the minimum of its potential, the anisotropy 
disappears. The question of when the anisotropy disappears depends on 
the parameter $m/\mu$ and the initial amplitude of $X$. The 
complete analysis of the initial 
conditions is possible as is done in the case of pure vector 
models~\cite{Chiba:2008eh}.

\begin{figure}
\includegraphics[height=8cm, width=13cm]{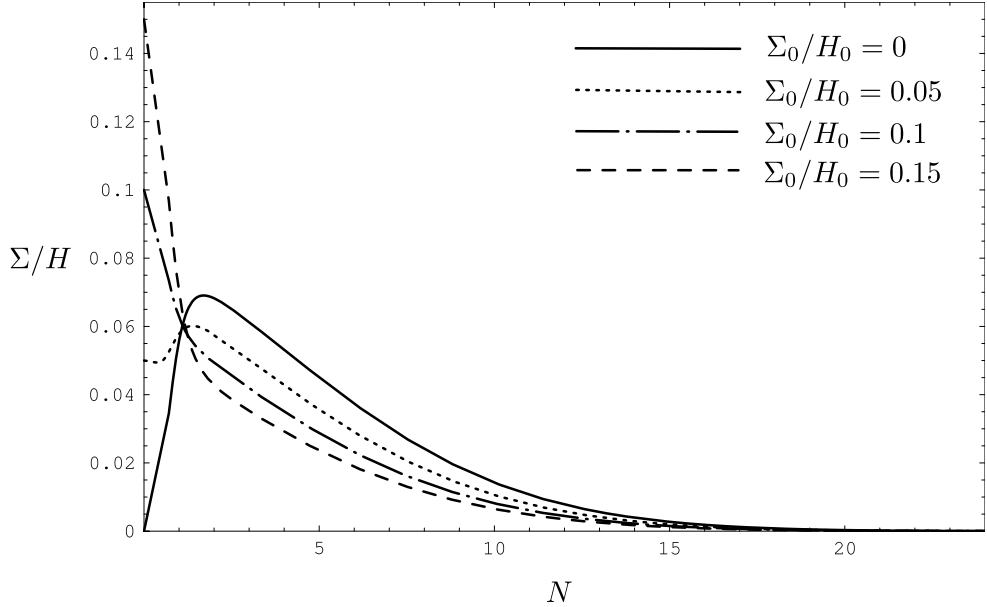}
\caption{The ratio $\Sigma/H$ is plotted as a function of e-folding number.
In spite of the rapid expansion of universe, the anisotropy
remains sizable for some period relevant to observations.}
\label{fig:2-b}
\end{figure}

 In Fig.~\ref{fig:1-b}, the phase flow in $\phi$ - $\dot{\phi}$ is plotted
with the same initial conditions as Fig.~\ref{fig:1-a}. 
The trajectories are almost same irrespective of the initial conditions for $X$,
and show the slow-roll phase.
 After slow-rolling, the inflaton field gets its damped oscillations.
 Therefore, the anisotropic inflation ends with reheating as in the
 standard inflation.

 In Fig.~\ref{fig:2-a}, we have plotted $H$ as 
 a function of e-folding number $N\equiv\alpha-\alpha_0$. 
Actually, we plotted four lines for different initial 
conditions $\Sigma_0/H_0=0,~0.05,~0.1,~0.15$ 
with fixed initial conditions $X_0=0.5$ 
and $\dot{X}_0=1.5\times10^{-5}$. In the Fig.~\ref{fig:2-a},
all lines are degenerated.
 Irrespective of the initial conditions, 
 we see the slow-roll phase from this figure.
As well as Fig.~\ref{fig:1-b}, qualitative behavior does not depend on the 
initial conditions for $X$ as long as its amplitude is small. 

%In the slow-roll phase, we find the anisotropy
%$\Sigma$ becomes $\kappa^2 X^2 H/3$ for small amplitudes
%$\kappa^2 X^2 \ll 1$ and $2 H/5$ for large amplitudes 
%from the formula (\ref{attractor}). In any case, the anisotropy 
%$\Sigma/H$ does not exceed $1$. 

 In Fig.\ref{fig:2-b},
we have plotted 
$\Sigma/H$ as a function of the e-folding number
with the same initial conditions as Fig.~\ref{fig:2-a}.
We see the anisotropy remains sizable for some period 
and then dumped down to zero
around $N\sim 20$ irrespective of its initial condition.
Duration of this phase depends on the mass of the vector field and other parameters.
We find larger initial condition for $\Sigma$ makes smaller anisotropy later. 
This can be understood from Eqs.~(\ref{eq:X}) and (\ref{attractor}).
Suppose the initial anisotropy is larger than the value calculated
from Eq.~(\ref{attractor}) using the initial data for $X, \dot{X}, H$,
then the anisotropy has to decrease rapidly to adjust to the value of 
the attractor.
During this phase, the effective mass in Eq.~(\ref{eq:X}) becomes large due to
the term $- \ddot{\sigma}_+$. Then, the amplitude of the vector $X$ 
before entering the slow-roll phase decreases
more than the case of smaller initial condition. From Eq.~(\ref{attractor}),
the small amplitude of $X$ implies
the small anisotropy during the slow-roll phase. 
Thus, we have understood  the behavior in Fig.~\ref{fig:2-b}.
In the case of the smaller initial anisotropy, the opposite happens. 

Previously, the conventional inflationary scenario in Bianchi type-I model
has been considered~\cite{Pereira:2007yy,Gumrukcuoglu:2007bx,Pitrou:2008gk}.
In those cases, the anisotropy decays exponentially fast.
In the recent  works~\cite{Koivisto:2008ig,Koivisto:2008xf},
the anisotropic universe has been investigated in the context of dark energy.
It would be interesting to apply our framework to the dark energy problem.

\section{Slow-roll approximation}
Now we consider the regime where
both of the vector and the scalar field are slow-rolling.
Thanks to the slow-roll for the vector field which 
breaks strong energy condition, we have enough anisotropic era.
For the reason explained in the previous section, we can exclude
$\sigma_{-}$ from the consideration. 
In the slow-roll approximation, we can write Eqs.~(\ref{eq:N})-(\ref{eq:X})
and (\ref{eq:sigma+})-(\ref{eq:alpha}) as
\begin{eqnarray}
&&3(-H^2+\Sigma^2)+\kappa^2V+2\kappa^2\Sigma^2X^2+\frac{\kappa^2}{2}\left(
m^2+\Sigma^2-4H\Sigma\right)X^2=0\,,
\label{eq:N2}\\
&&3H\dot{\phi}+V_{,\phi}=0\,,
\label{eq:phi2}\\
&&3H\dot{X}+\left(m^2-2H\Sigma-5\Sigma^2\right)X=0\,,
\label{eq:X2}\\
&&\Sigma =\frac{2X^2}{6/\kappa^2+5X^2}H\,,
\label{eq:sigma+2}\\
&&3H^2=\kappa^2V+\frac{\kappa^2}{2}\left(
m^2-2H\Sigma \right)X^2\,,
\label{eq:alpha2}
\end{eqnarray}
where we have ignored all second derivatives with 
respect to $t$ and square terms of the first derivatives with
respect to $t$ except for $H^2$, $\Sigma^2$ and $H\Sigma$. 
Since we are interested in the situation that the anisotropy is larger than
the slow-roll parameter, we have kept the anisotropy $\Sigma^2$ in the above equations. 
Adding Eqs.~(\ref{eq:N2}) and (\ref{eq:alpha2}), we obtain
Eq.~(\ref{eq:sigma+2}). Hence, Eq.~(\ref{eq:N2}) is redundant.

From Eq.~(\ref{eq:sigma+2}), we see $X$ should be small in order
for the anisotropy to be small as is mentioned at the end of the section II. 
Eliminating $\Sigma$ from Eqs.~(\ref{eq:sigma+2}) and (\ref{eq:alpha2}),
 we can deduce the Friedmann equation for this model
\begin{eqnarray}
H^2=\frac{6/\kappa^2+5X^2}{(6/\kappa^2+X^2)(3/\kappa^2+2X^2)}\left(
V+\frac{m^2}{2}X^2\right)\,.
\label{H}
\end{eqnarray}
Note that if there is no scalar field, i.e. no potential,
 this equation tells us that
$H\lesssim m$, no matter what value the vector field takes. This
contradicts the condition for the slow-roll: $H\gg m$. Thus
we find that the inflation does not occur only with the vector field.

From Eq.~(\ref{eq:sigma+2}) and (\ref{H}), we find
\begin{eqnarray}
\Sigma = \frac{2X^2}{\sqrt{(6/\kappa^2+5X^2)(6/\kappa^2+X^2)(3/\kappa^2+2X^2)}}
\sqrt{V+\frac{m^2}{2}X^2}  \ .
\label{S}
\end{eqnarray}
Here we have taken the positive sign because $H>0$.
 Inserting these results into Eqs.~(\ref{eq:phi2}) and (\ref{eq:X2}),
  we obtain a set of differential equations which determines $\phi$ and $X$.

Before proceeding to next section, we would like to clarify the scenario 
we envisage here. 
We are supposing the existence of inflaton and vector 
impurity.
The inflation is driven by the inflaton. 
In the case of chaotic inflation,
initial conditions for the inflaton are arbitrary as long as the energy
of the inflaton does not exceed the Planck energy. 
The larger initial amplitude for the inflaton gives the longer 
inflationary period. One of the initial conditions should correspond
 to our present universe.
Similarly, initial conditions for the vector field are arbitrary.
However, as we have seen in section II, the large amplitude of $X$ induces 
the tachyonic instability. Hence, this does not correspond to our universe.
The criterion for the stability condition $2H\Sigma \ll m^2$ 
in the slow-roll phase gives 
\begin{eqnarray}
\kappa \phi \ll \frac{m}{\mu} \frac{3}{\kappa X} \ .
\end{eqnarray}
where we used $\Sigma/H\sim\kappa^2X^2/3$ and $H^2\sim\kappa^2\mu^2\phi^2/6$
from Eqs.~(\ref{eq:sigma+2}) and (\ref{eq:alpha2}). 
We find that the inflaton field has the upper limit but
still has enough parameter range for the anisotropic
inflationary universe. For example, 
the parameters such as $\kappa\phi_0=10$, $\kappa X_0=0.5$ and 
$m/\mu \sim 2\sqrt{2}$,
which we used in the numerical calculations, satisfy this condition.

%===============================================================%
%************************ SECTION III ***************************%
%===============================================================%
\section{Implications for Primordial Fluctuations}

Although mixing between tensor and
scalar perturbations in anisotropic spacetime has been studied
in the literature (see, for example,~\cite{Pitrou:2008gk}),
there has been no concrete realization in the context of 
inflationary scenario. In this section, 
we would like to refresh various effects with our concrete
 anisotropic inflation model.
Here, we do not show the explicit numbers. However,
from the structure of the background equations, we guess possible
important consequences of vector impurity. 
Especially, we stress  the 
gravitational wave production due to the mixing 
 in the context of low scale inflation. 
Detailed calculations will appear in the follow-up paper
~\cite{kimura}.

Since the expansion is anisotropic, 
we expect statistically anisotropic
fluctuations. In fact, in the slow-roll phase 
($\dot{H},\dot{\Sigma}\simeq$ const.) , we have the metric
\begin{eqnarray}
  ds^2 = -dt^2 + e^{2Ht} \left[ e^{-4\Sigma t} dx^2 
           + e^{2\Sigma t} \left( dy^2 + dz^2 \right)\right] \ .
\end{eqnarray}
Now, let us consider a test scalar field $\psi$
in this background metric. According to Ref.~\cite{Ackerman:2007nb}, 
deviations from isotropy come in to the power spectrum of the form:
\begin{eqnarray}
  P_{\psi} ({\bf k}) 
  = P_0 (k) \left[1 + {\cal O} (\kappa^2 X^2) (\hat{\bf k}\cdot {\bf n})^2
  \right] \ ,
  \label{spectrum}
\end{eqnarray}
where $P_{\psi}$ is the power spectrum of the scalar field $\psi$,
$P_0 (k)$ is its isotropic part, ${\bf n}$ is the unit vector 
in the direction of $x$ and $\hat{\bf k}$ is the unit vector along the direction of
 wave number vector ${\bf k}$. Here, in evaluating the deviations, we used 
the approximate relation $\Sigma /H \sim \kappa^2 X^2$ derived from 
Eq.~(\ref{eq:sigma+2}).
Hence, if $P_0 (k)$ has a flat spectrum, the total spectrum should also 
be flat, even though it is anisotropic.
In the case of curvaton scenario, the scalar field $\psi$ decays after reheating
and converts to curvature perturbations. 
For this case, the fluctuation of $\psi$  gives CMB
fluctuations directly. Even for more complicated case of ours where 
inflaton also fluctuates,
we expect the similar result as Eq.~(\ref{spectrum}). This spectrum yields 
the anomaly in CMB observations.
In paper~\cite{Pullen:2007tu}, it is pointed out that about 10$\%$ deviation from the
isotropic statistic can be detected by WMAP and
2$\%$ by PLANCK. 
Hence, the interesting number is around ${\cal O} (\kappa^2 X^2) \sim 0.1$.

In addition to this apparent effect, we can expect much more interesting ones.
First of all, we should recall
 that the primordial gravitational waves are created quantum mechanically
 in the standard inflationary scenario.
Hence, their amplitude should be of the order of $H/M_{pl}$ ($M_{pl}$ is
the Planck mass).
On the other hand, here, we have another new component of 
tensor fluctuations induced by curvature perturbations
through the anisotropic background expansion. 
This can be understood from the background equation
(\ref{eq:sigma+2}):
\begin{eqnarray}
\frac{d}{dt}\left[ \sigma_{+} 
- \frac{1}{3} \kappa^2 X^2 \alpha  \right] =0  \ ,
\label{key}
\end{eqnarray}
where we have used the fact $\kappa^2 X^2 \ll 1$ and assumed
$X$ is almost constant. If we define the following quantity 
which is supposed to be the gravitational wave:
\begin{eqnarray}
  H_{+} = \delta \sigma_{+} 
- \frac{1}{3} \kappa^2 X^2 \delta \alpha\,,
\end{eqnarray}
we find this is a gauge invariant quantity in the long wavelength 
limit.
The background equation (\ref{key}) implies $H_{+}$
is a conserved quantity, namely, $dH_{+} /dt =0$. 
In the conventional isotropic case, this equation
reduces to $d \delta \sigma_{+} /dt=0$, which is equivalent to the
conservation law for gravitational waves $\dot{h}^{+} = 0$
in the long wavelength limit.
Here,  the polarized component of the transverse traceless tensor,  $h^{+}$,
goes to $\delta\sigma_+$ in the long wavelength limit, that is,
$h^{+} \stackrel{k\rightarrow 0}{\longrightarrow}\delta\sigma_+$.
Now, we have an extra term $\delta\alpha$ in 
the gauge invariant gravitational waves, 
so $\delta\sigma_{+}$ itself does not conserve in the
long wavelength limit. Since the curvature perturbation is dominant
during the inflation, the gauge invariant gravitational waves are 
generated first through $\delta\alpha$ which is supposed to be the
curvature perturbation. 
Thus, we have the relation $H_{+} = - \frac{1}{3} \kappa^2 X^2 \delta \alpha$.
As the vector field goes to zero, $\delta\sigma_+$
starts to appear because of the conservation law. Finally
$\delta\sigma_+$ becomes dominant after $X$ vanishes.
Thus, we have the relation $H_{+} = \delta \sigma_{+}$.
Because of the conservation law, we have 
$\delta \sigma_{+}|_{f} = - \frac{1}{3} \kappa^2 X^2 \delta \alpha |_{k}$,
where the left hand side is evaluated at the final time and the right hand side
 is evaluated at the horizon crossing time for the mode with wave number $k$.
 Now, let us define the curvature perturbations,  ${\cal R}$,
which goes to $\delta\alpha$ in the long wavelength limit, that is,
${\cal R} \stackrel{k\rightarrow 0}{\longrightarrow} \delta\alpha $.
Then, we obtain the final result
\begin{eqnarray}
  h^{+} = {\cal O}(\kappa^2 X^2 ) {\cal R} \ ,
  \label{main}
\end{eqnarray}
where the right hand side of Eq.~(\ref{main})
 are evaluated at the horizon crossing time. 
For the other polarization mode, we have the standard conservation law
$\dot{h}^{-} = 0$ in the long wavelength limit.
Here, we have the correspondence
 $h^{-}\stackrel{k\rightarrow 0}{\longrightarrow}\delta\sigma_-$.

The above discussion
 tells us the scalar perturbation ${\cal R}$, which comes from the fluctuations of
the trace part of metric (\ref{metric}), induces the tensor perturbation 
$h^{+}$, which is the one of the traceless part of the metric.  
 Therefore, the tensor power spectrum $P_{h^{+}}$ should be related to
  the curvature power spectrum $P_{\cal R}$ as
\begin{eqnarray}
  P_{h^{+}} = {\cal O} (\kappa^4 X^4 ) P_{{\cal R}} 
\end{eqnarray}
where the relation (\ref{main}) is used.
The other power spectrum $P_{h^{-}}$ coming from $\delta \sigma_-$ 
has no correlation with the curvature perturbations because there
is no conversion from the curvature perturbations in this case. 
In other words, we expect linear polarization in the primordial 
gravitational waves. 
This could be detected through CMB
or directly by DECIGO~\cite{Seto:2001qf}.

In papers~\cite{tomita,Matarrese:1997ay,Noh:2004bc,Osano:2006ew,Ananda:2006af,Baumann:2007zm},
the gravitational waves generated from the second order perturbations
have been studied. 
It should be stressed that the mechanism to produce the gravitational
waves we are discussing is the first order effect 
in the anisotropic inflationary universe.
The primordial fluctuations produced by the anisotropy
are not the conventional ones created from quantum fluctuations directly.
They are induced from the curvature perturbations created from quantum
fluctuations. Since the scalar perturbations are always
larger than the tensor perturbations, this anisotropy induced mechanism
is very efficient. 
This is an extremely important result. Because this implies
the existence of the primordial gravitational waves even in
low scale inflationary scenarios. 
In fact, supposing that we detect the anisotropy $\kappa^2 X^2 \sim 0.1$,
we would detect the primordial gravitational waves with the tensor-scalar
ratio $r\sim 0.01$.

Since at least part of the tensor perturbations are induced from the
curvature perturbations, 
there exists the correlation between the curvature and the 
tensor perturbations 
\begin{eqnarray}
  \left< h^{+} {\cal R} \right> 
  \sim {\cal O} (\kappa^2 X^2) 
  \left< {\cal R} {\cal R} \right>  \ ,
\end{eqnarray}
where  we used the relation (\ref{main}).
The correlation should give non-zero TB correlation at the 10$\%$ level
in the case of $\kappa^2 X^2 \sim 0.1$. In other words,
 the normalized correlation function 
$
  \frac{ \left< TB \right>}{\sqrt{\left< TT\right> \left< BB\right>}} 
  = {\cal O} (1)  
$
does not vanish. 
Surely, this should be searched observationally.

Although it is necessary to calculate precise numbers for prediction,
we have uncovered the new possibility to generate the
primordial gravitational waves on dimensional grounds
based on the concrete anisotropic inflationary model.
All of the above set of predictions are peculiar to the 
anisotropic inflation and hence can be regarded as signatures
of the vector impurity.

%===============================================================%
%************************ SECTION IV ***************************%
%===============================================================%
\section{Conclusion}

We have explored the cosmological implications of the vector impurity. 
It turned out that the vector impurity affects the cosmic no-hair 
theorem. Consequently, the anisotropic accelerating universe is possible
in the presence of the vector impurity. To prove it,
we have numerically solved the equations and shown the phase flow
from which we see the anisotropic inflation successfully occurs and ends with
oscillation. We found the explicit formula to determine the 
anisotropy of the inflationary universe by using the slow-roll 
approximation.

We have also discussed possible consequences of the
anisotropic inflation on the cosmological fluctuations.
Since rotational invariance is violated, the statistical
isotropy of CMB temperature fluctuations can not be expected.
It is intriguing to seek for the relation to the large scale
anomaly discovered in CMB 
by WMAP~\cite{Eriksen:2003db,Land:2005ad,Copi:2005ff}.
More interestingly, 
the tensor perturbations could be induced from the
curvature perturbations through the anisotropy of the background
spacetime. One immediate consequence is the correlation between 
the curvature perturbations and the tensor perturbations.
This correlation should be detected through the analysis of 
temperature-B-mode correlation in CMB. 
This new possibility implies, even in the low scale inflation, we can expect
the primordial gravitational waves. This is an important result
for future observational planning, 
because there has been worry that string cosmology tend to suggest
the low scale inflation. Moreover, 
because of the anisotropy, there might be linear polarization
in the primordial gravitational waves.  
This polarization can be detected either through the CMB observations
or direct interferometer observations. 
These predictions can be checked by future observations.

To make these predictions more precise, we need to develop
the  perturbative analysis~
\cite{Pereira:2007yy, Pitrou:2008gk, Gumrukcuoglu:2007bx,Tomita:1985me}. 
These are now under investigation~\cite{kimura}.
The calculation of the perturbations is much more complicated
due to the violation of rotational invariance.
However, since the anisotropic inflationary universe
is smoothly connected to the isotropic radiation dominant phase,
the interpretation of the results should be clear. 
The implications in primordial magnetic fields and the structure formation 
of the universe~\cite{Ando:2008zz} should be also studied in future work. 

\begin{acknowledgements}
We are grateful to Misao Sasaki for detailed discussions, and for
emphasizing the importance of gauge invariant analysis.
We wish to thank Takeshi Chiba, Hideo Kodama, Shinji Mukohyama,
and Naoshi Sugiyama for fruitful discussions. 
SK is supported by World Premier International Research Center
Initiative (WPI Initiative), MEXT, Japan.
MK is supported by a Grant-in-Aid for JSPS Fellows. 
JS is supported by the Japan-U.K. Research Cooperative Program, 
Grant-in-Aid for  Scientific Research Fund of the Ministry of 
Education, Science and Culture of Japan No.18540262 and No.17340075. 
SY is supported in part by Grant-in-Aid for Scientific Research
on Priority Areas No. 467 ``Probing the Dark Energy through an
Extremely Wide and Deep Survey with Subaru Telescope'', by the
Mitsubishi Foundation, and by Japan Society for Promotion of Science 
(JSPS) Core-to-Core Program ``International Research Network for 
Dark Energy'', and by Grant-in-Aids for Scientific Research
(Nos.~18740132,~18540277,~18654047). 
\end{acknowledgements}

\end{document}